\documentclass[journal=jacsat,manuscript=article]{achemso}

\usepackage[version=3]{mhchem} 
\usepackage[utf8]{inputenc}
\usepackage[T1]{fontenc}
\usepackage{graphicx}
\usepackage{amsmath}

\usepackage{xcolor}
\usepackage{ulem}

\title{The true circular dichroism of optically active achiral metasurfaces and its relation with chiral nearfields}

\author{Mathieu Nicolas}
\affiliation{Sorbonne Universit\'{e}, CNRS, Institut des NanoSciences de Paris, INSP,  F-75252 Paris, France}
\alsoaffiliation{Sorbonne Universit\'{e}, CNRS, Laboratoire de R\'{e}activit\'{e} de Surface, LRS,  F-75252 Paris, France}
\author{Per Magnus Walmsness} 
\affiliation{Department of Physics, Norwegian University of Science and Technology (NTNU), Trondheim, N-7491, Norway} 
\author{Jayeeta Amboli}
\affiliation{Aix-Marseille Univ, CNRS, Centrale Marseille, Institut Fresnel 13013 Marseille, France}
\author{Lu Zhang}
\affiliation{Sorbonne Universit\'{e}, CNRS, Institut des NanoSciences de Paris, INSP,  F-75252 Paris, France}
\alsoaffiliation{Sorbonne Universit\'{e}, CNRS, Laboratoire de R\'{e}activit\'{e} de Surface, LRS,  F-75252 Paris, France}
\author{Guillaume Demesy}
\affiliation{Aix-Marseille Univ, CNRS, Centrale Marseille, Institut Fresnel 13013 Marseille, France}
\author{Nicolas Bonod}
\affiliation{Aix-Marseille Univ, CNRS, Centrale Marseille, Institut Fresnel 13013 Marseille, France}
\author{Souhir Boujday}
\affiliation{Sorbonne Universit\'{e}, CNRS, Laboratoire de R\'{e}activit\'{e} de Surface, LRS,  F-75252 Paris, France}
\author{Morten Kildemo} 
\affiliation{Department of Physics, Norwegian University of Science and Technology (NTNU), Trondheim, N-7491, Norway} 
\author{Bruno Gallas}
\affiliation{Sorbonne Universit\'{e}, CNRS, Institut des NanoSciences de Paris, INSP,  F-75252 Paris, France}
\email{bruno.gallas@sorbonne-universite.fr}

\begin{document}

\begin{abstract}
\textit{Optically active achiral metasurfaces offer a promising way to detect chiral molecules based on chiroptic methods. The combination of plasmonic enhanced circular dichroism and reversible optical activity would boost the sensitivity and provide enantiomer-selective surfaces while using a single sensing site. In this work, we use metasurfaces containing arrays of U-shaped resonators as a benchmark for analyzing the optical activity of achiral materials. Although the peculiar optical activity of these metasurfaces has been quite well described, we present here an experimental and numerical quantitative determination of the different contributions to the measured optical activity. In particular, it is shown that linear birefringence and retardance contribute, but only marginally, to the apparent circular dichroism of the metasurface associated with the excitation of magnetoelectric modes. We then numerically demonstrate the peculiar near-field properties of the magneto-electric modes and explain how these properties could be reflected in the far-field polarimetric properties in the presence of chiral molecules. This work provides alternatives for the detection scheme of chiral molecules using plasmonic resonators. 
}
\end{abstract}
\section{Keywords} 
Circular Dichroism; Mueller matrix; Plasmonics; magneto-electric coupling; Differential decomposition
\section{Introduction}

The manipulation of polarimetric properties of light using surfaces has been intensively studied for many years due to the strong effects that can be obtained by using anisotropic plasmonic resonators arranged in metasurfaces. An important polarimetric property is circular dichroism (CD), which finds many applications in sensing of chiral molecules \cite{Kadodwala2010,Naik2010,Markovich2013,Kadodwala2015,Quidant2018,Quidant2020}, lighting \cite{Kivshar2022}, light steering \cite{Capasso2011,Capasso2014}. Classically, the CD measured in the far-field originates from the absorption difference between right and left circularly polarized light (RCP and LCP), also called Circular Differential Optical Absorption (CDOA), and is obtained by transmission measurements. Ideally, CDOA = CD, and it is recognized that CD is associated with chiral objects. To enhance the interaction of light with chiral molecules, chiral plasmonic resonators have been used.  \cite{Kadodwala2010,Kadodwala2015,Landobasa2017,Quidant2018}. However, recent numerical studies have suggested that observed CDOA may actually result from a combination of linear birefringence and dichroism rather than true CD \cite{Peak2009,Arteaga2016,Orrit2022}. It was also shown that sample imperfections arising during fabrication may also make the nanostructures at the surface actually 3D~\cite{Arteaga2016}, or even that polarization mixing originating in spatial dispersion should yield apparent CDOA \cite{Gompf2011,Guth2012}. Optically active achiral resonators can be used to minimize spurious effects due to fabrication imperfections that prevent direct comparison of the CD of metasurfaces with opposite plasmonic enantiomers. Optically active achiral resonators are based on the excitation of electric and magnetic dipolar moments with magneto-electric coupling and were first introduced as pseudo-chiral resonators in Ref.\cite{Engheta92}. Owing to the magneto-electric coupling, the polarizability tensor is bi(an)isotropic (see Ref. \cite{Sturm2021} for instance for a comprehensive review) resulting in the possibility of probing both CDOA on the same metasurface \cite{Rockstuhl2007,Proust2016,Koenderink2009,Guth2012}. In addition to measuring the far-field CDOA, metasurfaces for sensing must also generate chiral fields in their vicinity, and therefore the relationship between the near-field chiral field and the far-field CDOA must be carefully investigated. \\
In this work, we experimentally determine the polarimetric response of metasurfaces containing optically active achiral resonators, using the differential decomposition to identify the different origins of the CDOA with respect to the resonant modes excited in the resonators. We then numerically study the near-field polarization properties of the resonators with emphasis on chirality, and discuss the potential relationships between the far-field CD, the presence of local chiral fields, and how these properties could be used for sensing applications.

\section{Experiments}
\subsection{Manufacture U-shaped resonators}
\begin{figure}[htbp]
\centering\includegraphics[width=8.5 cm]{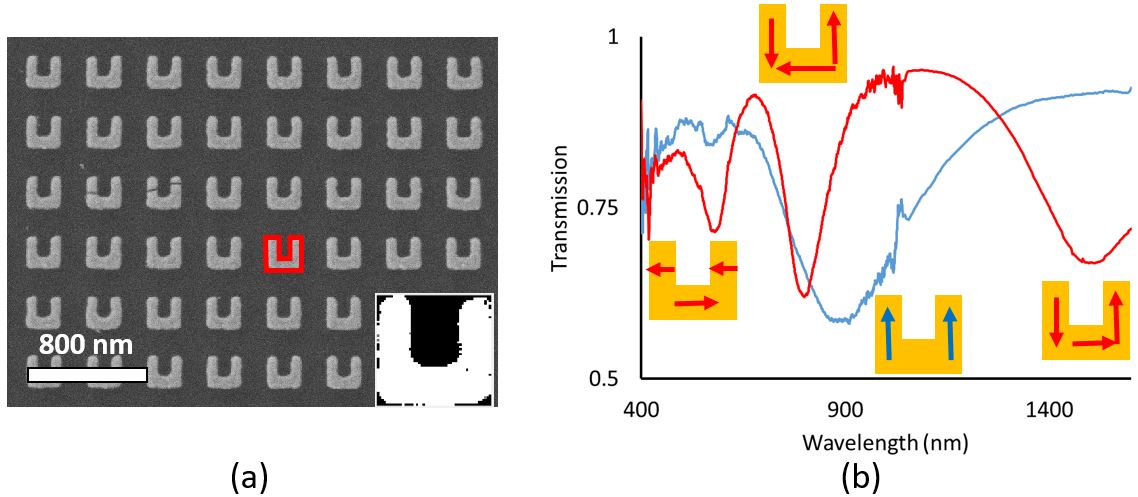}
\caption{(a) SEM image of a metasurface. The red line delineates the ideal shape expected, the inset presents the average shape of a resonator extracted from image analysis. (b) Transmission measurements performed at normal incidence on the metasurface for polarizations along (red line) and perpendicular (blue line) to the bottom arm of the resonators. The arrows in the schematics of the resonators show the instantaneous current distributions in the resonators at the resonances. }
\label{fig:SEM}
\end{figure}

We have used metasurfaces made of arrays of U-shaped resonators. The metasurfaces were obtained using a conventional e-beam lithography process. Clean silica substrates (fused SiO$_2$) were covered with 120 nm PMMA and 20 nm of conductive resist (electra92) to evacuate the charges during the e-beam writing. After writing the samples were revealed and then a 4 nm titanium wetting layer followed by a 40 nm thick layer of gold were deposited. A lift-off process was finally performed to reveal the U-shaped resonators. 
It can be seen in Figure \ref{fig:SEM}(a) that the metasurface contains resonators exhibiting a fairly good shape homogeneity although close inspection shows that each resonator is slightly different from its neighbour. This is typically a result of the polycrystallinity of the evaporated Au film. Small systematic deviations were evaluated from the analysis of the SEM image. The inset in Figure \ref{fig:SEM}(a) shows the shape of the average U-shaped resonator which exhibits some slight systematic asymmetry. 

\subsection{Normal incidence transmission measurements}

The transmission measurements at normal incidence were made in an optical microscope (Olympus) with a $\times 10$ objective. The light was first filtered with a linear polarizer aligned either parallel to or perpendicular to the bottom arms of the resonators. The metasurface was imaged on the entrance of a fiber connected to two CCD spetrophotometers (400-1000 nm and 1000-1700 nm). The intensity transmitted through the metasurface was then normalized to the intensity transmitted through the glass. \\

\subsection{Transmission Mueller Matrix Ellipsometry}
The polarimetric properties of the metasurfaces were measured using a spectroscopic ellipsometer (RC2 JA Woolam Company) fitted with focusing probes (numerical aperture 0.017). The samples were placed on a rotating sample holder and aligned with respect to the mechanical axes of the ellipsometer. Measurements of the full transmission Mueller matrix were performed in transmission in the 400-1600 nm range at a polar angle of incidence of 15$^o$. The sample was rotated by 360$^o$ around the sample normal and thereby measurements were performed with azimuthal angle of incidence every 2.5$^o$. \cite{Brakstad2015} \\

\subsection{Numerical simulation details}
Simulations were performed using Finite Element Methods (home-made code, see Supporting Information S.I.1). The U-shaped resonators were located on a periodic lattice of pitch 400 nm. The dimensions of the arms were extracted from the SEM images and the thickness was set to 40 nm. The optical constants of gold were obtained by keeping the interband transitions of bulk gold and adding a Drude term damped by a factor 3 to reproduce the effect of scattering at the surface or at grain boundaries. The Ti wetting layer was not included in these simulations. The optical constants of the substrate were those of silica. The polarization properties obtained in the far field were decomposed on a Jones vector, then transformed in Mueller matrix elements using the Jones-Mueller transformation~\cite{Walmsness2019}. 

\section{Polarimetric properties}
\subsection{Normal incidence transmission}
The experimental Transmission for linear polarization along and perpendicular to the bottom arm of the resonator performed on the metasurface are shown in Figure \ref{fig:SEM}(b). Different dips in the transmission spectra, associated with absorption bands, were observed. For light polarized along the bottom arm of the resonators (red line), they were located at about 1500, 820 and 580 nm. For light polarized perpendicular to the bottom arm (blue line), one main absorption band was observed at 900 nm and a smaller one at 585 nm. 
With the average dimensions extracted from the SEM images and same incident polarizations, the calculated transmissions showed strong dips at 1630 nm, 805 nm and 580 nm for horizontal polarization and at 840 nm for vertical polarization (Supplementary Information S.I.1). These values were in qualitative good agreement with the measured ones. The differences between calculations and measurements originated most likely from the differences between the ideal shape used in the model and the fabricated one and in the absence of the Ti adhesion layer in the simulation model. 
The current distribution calculated numerically in the resonators at the resonance wavelength are shown as insets in Figure \ref{fig:SEM}(b) to get better insight into these resonances. It can be seen that the resonances at 1500 nm and 820 nm are characterized by currents flowing in opposite directions in the lateral arms of the resonators. These modes are characterized by a magneto-electric coupling which induces measured CD in the far-field for non-normal incidence \cite{Proust2016,Koenderink2009,Guth2012}. These modes are referred to the literature as magnetic modes. The observation of optical activity at oblique incidence associated with the magneto-electric coupling of the magnetic mode has been shown at the level of the single U-shaped resonator both numerically and experimentally \cite{Proust2016,Varault2013}. The mode excited for light perpendicular to the bottom arm is charaterized by currents flowing in phase in the lateral arms of the resonator and is referred to as electric mode. The resonance near 580 nm is associated with transverse modes of the arms and has a quadrupolar character \cite{Proust2016,Koenderink2009}. 

\subsection{Transmission Mueller Matrix with full azimuthal rotation}
\begin{figure}[htbp]
\centering\includegraphics[width=8.5 cm]{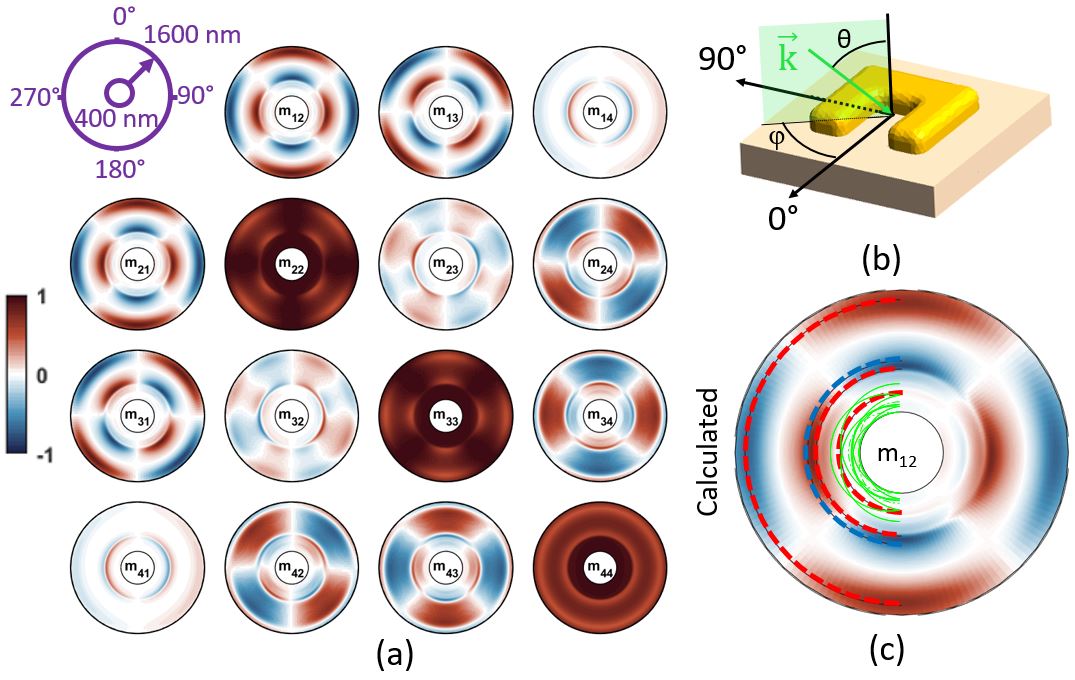}
\caption{(a) Measured Mueller matrices of the metasurface of Figure \ref{fig:SEM} as a function of azimuthal angle of incidence and wavelength (radial dimension). All elements are normalized to the first one which is not presented. The element m$_{11}$ was replaced by a schematic of the coordinate system used. (b) 3D view of the unit cell used for the numerical simulations with azimuthal ($\phi$) and polar ($\theta$) angles of incidence. For any given incident field direction, defined by its wavevector $\vec{k}$, the p-polarization was defined for a wave with its electric field in the plane of incidence (green plane), the s-polarization being perpendicular to the plane of incidence. (c) Element m$_{12}$ calculated using FEM. The blue dotted lines, resp. red, indicate the spectral positions of the electric modes, resp. magnetic. The location expected for the Rayleigh lines are plotted in green, dotted lines for the air modes and full lines for the substrate modes.}
\label{fig:MMmes}
\end{figure}

Figure \ref{fig:MMmes}(a) shows the measured Mueller matrix obtained from a selected metasurface. The radial scale is the wavelength and it ranges from 400 nm for the inner ring to 1600 nm at the outer ring ; in the plot the polar scale represents the azimuthal angle of incidence in the measurement. The azimuthal angles of incidences 90$^o$ and 270$^o$ correspond to light propagating in the plane of incidence containing the bottom arm of the resonators (Figure \ref{fig:MMmes}(b)). A very good agreement was obtained between the measured and the numerically calculated values. The calculated element m$_{12}$ is presented in Figure \ref{fig:MMmes}(c), the other elements are presented in Supplementary Information S.I.2.

The signature of the magneto-electric coupling is clearly visible in the values of the off-diagonal blocks of the Mueller matrix which are not vanishing for light propagating in the plane containing the bottom arms of the resonators (90$^o$ and 270$^o$). In particular the elements m$_{14}$ and m$_{41}$, related to circular dichroism optical absorption,\cite{LuChipman96} show large values for these directions of propagation and change sign when the azimuthal angle of incidence changes from 90$^o$ to 270$^o$ as can be seen in Figure \ref{fig:MMmes}(a)~\cite{Yoo2018}.
Calculations using finite element methods show that ideal samples exhibit the same polarimetric features as the measured ones as shown in Figure \ref{fig:MMmes}(c) (see Supplementary Information S.I.2 for the full calculated values) and that the deviations from ideal resonators as shown in Figure \ref{fig:SEM}(a) do not modify noticeably the properties of the metasurface. The relation between the excited modes and the Mueller matrix elements are presented in Figure \ref{fig:MMmes}(c). Conventionally, the element m$_{12}$ presents the linear dichroism. According to the transmission measurements, we expect to see maxima along the azimuthal angles of incidences 0$^o$,  90$^o$,  180$^o$ and  270$^o$. This is exactly what can be observed in Figure \ref{fig:MMmes}(c). We have also added dotted lines at the location of the modes observed in transmission, i.e. at 1500, 820 and 580 nm for horizontal polarization (marked as red dashed lines) and 900 nm for vertical polarization (marked as blue dashed lines). Recalling that \cite{Brakstad2015} 
$m_{12}=\left(|t_{\textup{pp}}|^2+|t_{\textup{sp}}|^2-|t_{\textup{ps}}|^2-|t_{\textup{ss}}|^2 \right)/\left(|t_{\textup{pp}}|^2+|t_{\textup{sp}}|^2+|t_{\textup{ps}}|^2+|t_{\textup{ss}}|^2 \right) $, where $t_{\rm{pp}}$,$t_{\rm{ss}}$ , $t_{\rm{sp}}$ and $t_{\rm{ps}}$ are the transmission amplitude coefficients, it is easily understood that the maximum of the linear dichroism is observed at these modes/wavelengths. Since the resonators are on a square lattice, the scattered light should propagate only in the directions associated with the diffracted orders. The onset of the excitation of diffracted orders yields sharp structures named Rayleigh anomalies in the Mueller matrix elements. The signature of the Rayleigh anomalies was also observed (the calculated Rayleigh lines are shown as full green lines (Air) and green dashed lines (glass) in Figure \ref{fig:MMmes}(c)), but it should be noted that the first anomaly is at 700 nm and it should not affect too much the line shape of the electric and magnetic modes, unlike the quadrupolar modes~\cite{Walmsness2019}. 

\begin{figure}[htbp]
\centering\includegraphics[width=8.5cm]{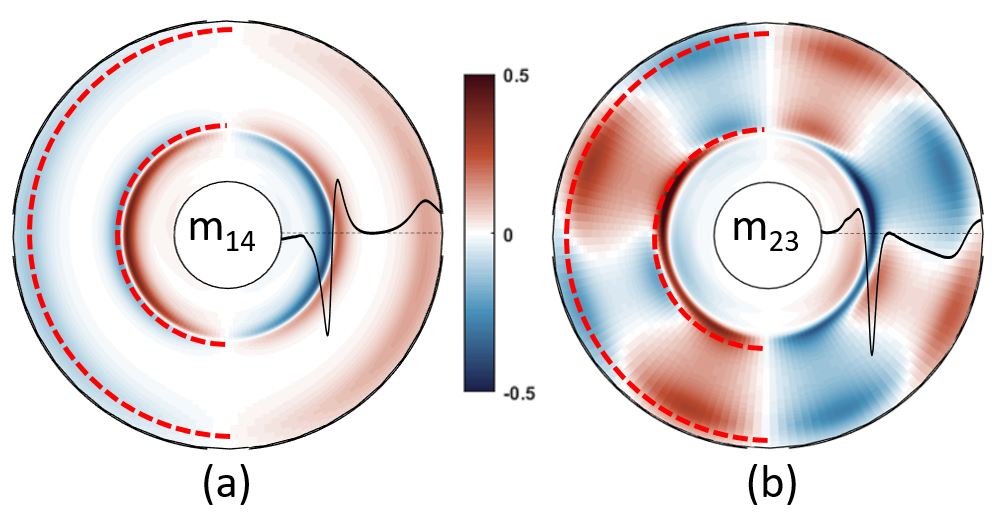}
\caption{(a) Measured element m$_{14}$ and (b) m$_{23}$. The red dotted circles indicate the position of the magnetic modes. The black curve shows the spectral variations of m$_{14}$ and m$_{23}$ for the azimuthal angle 90$^o$.}
\label{fig:LinPolAng}
\end{figure}

The elements m$_{14}$ and  m$_{23}$ are related to the amplitude and phase of the CDOA and should arise from the magneto electric coupling in optically active achiral metasurfaces illuminated at non-normal incidence. As already described in the literature \cite{Proust2016,Yoo2018}, it can be observed from Figure \ref{fig:LinPolAng}(a) that the spectral dependence of the element m$_{14}$ exhibited a characteristic S-shape crossing zero at the position of the magnetic mode at 815 nm which has the strongest magneto-electric contribution \cite{Proust2016,Koenderink2009}. This crossing was associated with an extremum in the element m$_{23}$ as can be observed in Figure \ref{fig:LinPolAng}(b). However, the maximum values in m$_{23}$ were not obtained at the azimuthal angles of incidence 90$^o$ and 270$^o$ as a result of the interplay between the magneto-electric coupling yielding the 2-fold symmetry as observed in Figure \ref{fig:LinPolAng}(a) and the linear polarization effects associated with the biaxial anisotropy of the resonators yielding the additional 4-fold symmetry (Figure \ref{fig:LinPolAng}(b) and S.I.2). A similar signature was observed in the spectral region of the magnetic mode at 1500 nm, but with smaller variation, owing to the weaker magneto-electric coupling associated with this mode \cite{Proust2016}. 

\begin{figure}[htbp]
\centering\includegraphics[width=10 cm]{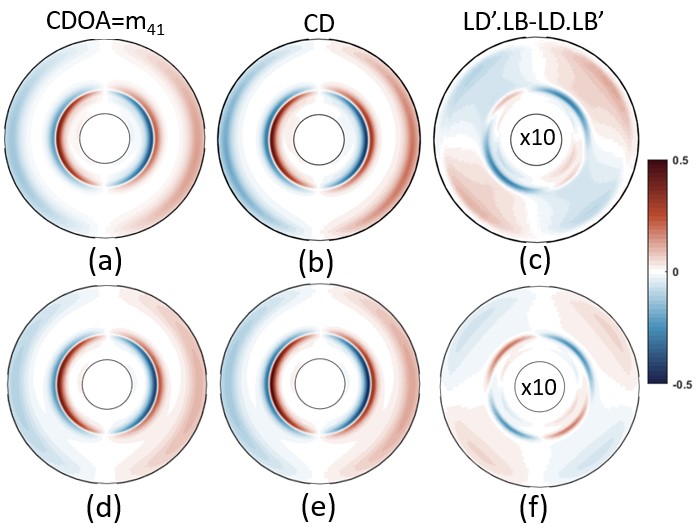}
\caption{(a) Measured element m$_{41}$ of the measured Mueller matrix, (b) Experimental CD as determined from the element (4,1) of the matrix L$_m$ and (c) Experimental contribution of the linear dichroism and birefringence to the CDOA. (d), (e) and (f) Numerically determined values of m$_{41}$, CD and contribution of linear effects to CDOA. All plots have the same colour scale but the values of (c) and (f) have been multiplied by a factor of 10 for a better comparison.}
\label{fig:CDmm41}
\end{figure}

To disentangle the contributions of circular dichroism and linear dichroism in the off-diagonal blocks of the Mueller matrix, the true polarimetric properties were obtained through the so-called polar decomposition \cite{Ossikovski2007,LuChipman96,Ossik2011,Arteaga2013}. In transmission configuration, provided that the sample is thin and homogeneous throughout its thickness, the differential decomposition is unique and provides the true polarimetric properties \cite{Ossikovski2007}. Figure \ref{fig:CDmm41} presents the comparison between the measured element CDAO=m$_{41}$ (Figure \ref{fig:CDmm41}(a)) and the CD as determined from the element (4,1) of the antisymetric part of the logarithm of the Mueller matrix L$_m$  (see Figure \ref{fig:CDmm41}(b) and also S.I.1, Eq. (4)). The contribution of the linear dichroism and birefringence to m$_{41}$ was obtained as 0.5(LD'.LB-LD.LB') and is plotted in Figure \ref{fig:CDmm41}(c).\cite{Peak2009} These latter values were multiplied by a factor of 10 to be visible on the same colour scale as m$_{41}$ and CD, but they still remained small. Figures \ref{fig:CDmm41}(d), (e) and (f) present the same analysis made from the numerically calculated values. The contribution of the linear polarization to the CDOA is definitely present, but is at least one order of magnitude smaller than CD. This evidences that although linear dichroism contributes to CDOA, the main origin of CDOA in optically active achiral metasurfaces lies in CD. The remaining linear polarization effects contributing to CDOA may partly originate from the unintended asymmetry of the resonators as shown in Figure \ref{fig:SEM}(a). However the same effects were observed in the numerically calculated polarimetric properties of perfectly symmetric resonators (Figure \ref{fig:CDmm41}). This was particularly visible at normal incidence (S.I.2) where residual CDOA was observed which could be attributed completely to linear polarization effects. The contribution of linear polarization effects was also particularly visible when comparing the elements m$_{23}$ and L$_m$(2,3) where the strong 8-fold symmetry associated with the biaxial anisotropy of the resonators was completely removed in L$_m$(2,3) (S.I.2). This last point stresses how important it is to not mistake CDOA for CD nor optical rotation for CB in metasurfaces. 

\section{Numerical results for the near-field}
In addition to manipulating the polarization properties in the far field, optically active achiral metasurfaces can also be used to manipulate the vectorial properties in the near field. Considering the use of such metasurfaces as sensing elements for the detection of chiral molecules, we are interested in normal incidence excitation, where the contribution of the bare metasurface to the measured far-field CD (elements m$_{14}$ and m$_{23}$) would be zero.\cite{Proust2016,Koenderink2009} We have used numerical simulations to evaluate the properties of the near-field, owing to the very good qualitative agreement between the measured and computed far-field properties in transmission presented here and in Supporting Information S.I.2. Different quantities can be used to characterize the local field : intensity, chirality density and ellipticity which are all time-averaged quantities.\cite{Nori2014,Nori2015} The intensity $\vert \rm{E} \vert^2$ is the main property used in sensing applications relying on plasmonics and is directly related to the existence of resonant modes. The ellipticity $\sigma_z$ is directly related to the polarization of the light. The chirality density C is involved in the differential absorption of circularly polarized light by chiral molecules \cite{TangCohen} and is defined as C=$-( \omega  \epsilon_o / 2).\rm{Im}(E^*.B)$. 

\begin{figure}[htbp]
\centering\includegraphics[width=8.5cm]{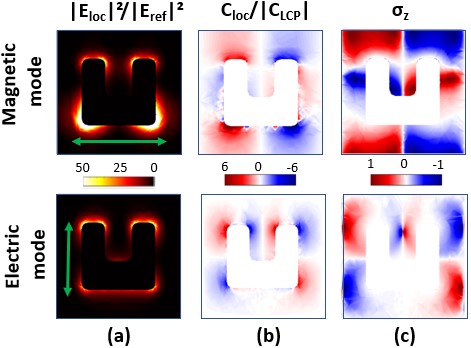}
\caption{Near-field properties calculated for excitation at normal incidence for the magnetic mode at 820 nm and for the electric mode at 900 nm. (a) Electric field enhancement, (b) Chirality density normalized to that of circularly polarized light, (c) Ellipticity $\sigma_z$. The green arrows in (a) indicate the incident polarization direction.}
\label{fig:nearfield}
\end{figure}

Figure \ref{fig:nearfield} presents the distributions of the field intensity, C and $\sigma_z$ calculated in the unit cell around one resonator at 20 nm above the glass substrate, i.e. in the middle plane of the U-shaped resonator. The chirality density was normalized to that of a circularly polarized light 20 nm above a bare substrate. The fields were calculated at 820 nm with light polarized along the bottom arm of the resonator for the magnetic mode and at 900 nm with light polarized perpendicular to the bottom arm for the electric mode. An incidence of 0$^o$ was used. It must be noted that at 15$^o$ of incidence and azimuthal angles of 90$^o$ and 270$^o$ very similar maps would be observed with a slight asymmetry in the field distributions (Supporting Information S.I.3). 

In the near-field of the resonators, both modes exhibit locally chiral light, although these properties average to zero over the whole unit cell. For the chirality density, the values are larger than unity (Figure \ref{fig:nearfield}(b)) because of the field enhancement (Figure \ref{fig:nearfield}(a)) and of the ellipticity of the local light (Figure \ref{fig:nearfield}(c))~\cite{Nori2014,Nori2015}. In contrast, the ellipticity $\sigma_z$ is normalized to the local intensity and is always comprised between -1 and 1 \cite{Andrews2012}. For a plane wave propagating along $z$, ellipticity and chirality density are related to each other and we would expect $C=\sigma_z \mid\vec{\rm{E}}\mid^2$.\cite{Nori2015} However, in the near field this relation breaks down, in particular in the close vicinity of the resonators (Supporting Information S.I.3)~\cite{Nori2015,Nechayev2019}. In Figure \ref{fig:nearfield}(c), it is interesting to notice that even if at 0$^o$ of incidence m$_{41}=0$ in the far-field, $\sigma_z \neq 0$ in the vicinity of the resonator and can locally reach values near unity characteristic of circular polarization. 
Beyond these observations common to the two modes, fundamental differences can be observed between the electric and magnetic modes. 
For the electric mode, the spatial distribution of the chirality density around each arm of the U-shaped resonators is mostly similar to what has been described in the case of isolated nanorods \cite{Okamoto2018} or for an electric dipole \cite{Giessen2012}, with a slight modification owing to the presence of the bottom arm. This is not surprising since the currents flow in phase in these arms (Figure \ref{fig:SEM}(b)). We observe a vertical mirror symmetry of the chiral properties at the level of each vertical arm. 
In the case of the magnetic mode, the properties associated with the chirality of light are much stronger than for the electric one. This results partly from the stronger value of the near-field (Figure \ref{fig:nearfield}(a)), but this was also observed for $\sigma_z$ (Figure \ref{fig:nearfield}(c)). The vertical mirror symmetry of the chiral properties is observed only at the scale of the full resonator. In particular, the chirality density shows one single sign at the end of each arm (Figure \ref{fig:nearfield}(b)). 

\begin{figure}[htbp]
\centering\includegraphics[width=8.5cm]{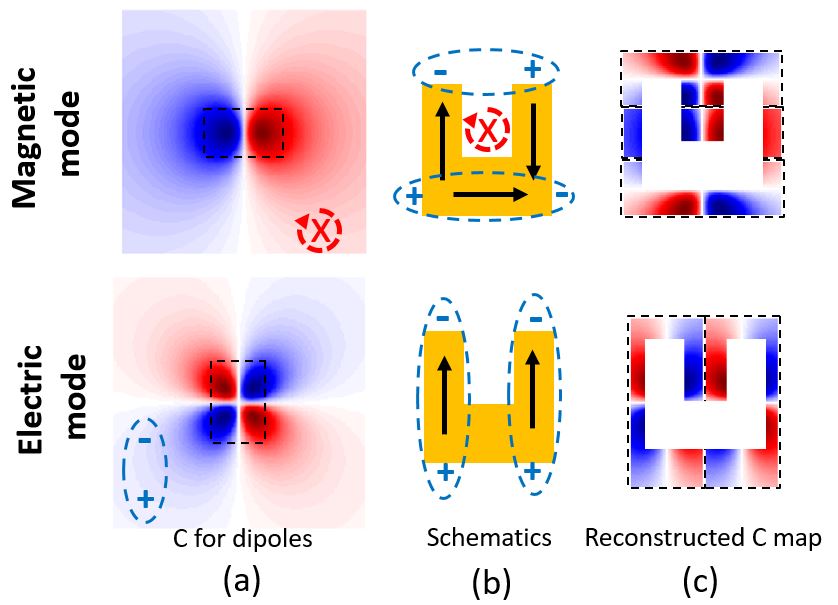}
\caption{(a) Spatial distribution of the chirality density near an electric dipole (blue dotted line) and a magnetic dipole (red dotted line). (b) Schematics of the current (black arrows) and charge distributions (blue signs) at the electric and magnetic modes of one U-shaped resonator. (c) Reconstruction of the chirality density distribution near the U-shaped resonator. The white resonator shows where the resonator could be withing the map.}
\label{fig:PLD_mod}
\end{figure}

These spatial distribution of chirality density can be reproduced qualitatively using a point-like dipole model taking into account both near and far fields (Figure \ref{fig:PLD_mod}(a)). We have calculated the fields at 20 nm above dipoles excited at normal incidence. For the magnetic mode, the magnetic dipole contribution was added thanks to its excitation through the magneto-electric coupling even at normal incidence. The current distributions and charge distributions were used to define the number and orientations of the electric and magnetic dipoles for the electric and magnetic modes (Figure \ref{fig:PLD_mod}(b)). We have assumed that the magnetic mode was excited by the interaction of the electric field of light with the bottom arm of the resonator. Owing to the very short distance between the different dipoles and the observation plane, we have neglected interference effects and have simply added the different contributions. 
It was possible to sketch the overall chirality density map in the vicinity of the resonator in figure \ref{fig:PLD_mod}(c) by superimposing the chirality density maps, arbitrarily selected in the black dotted line of figure \ref{fig:PLD_mod}(a), for the magnetic dipole and the two electric dipoles as shown in schematic (b). Note that, for illustration purpose, the colour scales were all normalized so that the spatial variations of the chiral fields within this model are not quantitative. However qualitative, the comparison of Figures \ref{fig:nearfield}(b) and Figures \ref{fig:PLD_mod}(c) showed a very good agreement between numerically calculated and modelled spatial distributions of the chirality density. This model gives then a good insight in the chiral properties of the light in the near-field from the knowledge of the modes excited.

From the perspective of using the enhanced chirality density of light in the vicinity of the resonators, with a detection in the far-field, it appears that the magnetic mode should be favoured. First, thanks to the largest values of chirality density observed in Figure \ref{fig:nearfield}(b). However, more important are probably the symmetries of the chiral fields around the resonators, assuming that the sensitivity to the presence of absorbed molecules will come from the end of the arms where both electric field and ellipticity are large. At the electric mode, the chirality density distribution presents mirror antisymmetry at the level of each arm's end. In the case of absorbed chiral medium, each arm's end would probe both chiral properties hence averaging the response to zero.\cite{Weis2022} 

For the magnetic mode, the left and right lateral arm's end would probe the opposite polarization properties of any absorbed chiral medium. This sensing effect would be happening even at normal incidence: one should expect the appearance of some mirror asymmetry in the near-field resulting in CDOA in the far-field polarimetric response of the metasurface coupled to chiral molecules.\cite{Markovich2013,Naik2010,Quidant2020,Weis2022} That should affect mostly the elements associated with linear dichroisms and birefringences instead of circular ones, most likely the elements mm$_{13}$ and mm$_{24}$. In addition, it has been reported that the local chirality density state might be better reported by other elements than CD \cite{Poulikakos2016}. This suggests that full polarimetric measurements would be mandatory to investigate separately the enhancement of the CD of the chiral molecules resulting merely from enhancement of the intensity of the nearfield and the influence of the chiral molecules on the plasmon resonances supporting modes with magneto-electric coupling.

\section{Conclusion}
In conclusion, from the measured Mueller matrix response of an optically active achiral metasurface composed of U-shaped resonators, we have shown through the differential decomposition that the true optical activity accounted to more that 90 $\%$ of the \textcolor{red}Circular Differential Optical Absorption (CDOA). These observations agree well with the numerically calculated values obtained for ideal samples, hence evidencing that linear birefringence and retardance also contribute to the CDOA. We show that the nearfield exhibits elliptically polarized light even for normal incidence illumination. Very different spatial symmetries at the electric and magnetic resonances are obtained which can be understood using a point-like dipole decomposition of the radiation pattern of the resonators. In particular, we show that the ellipticity and chirality densities exhibit opposite signs at the end of each arms which hints toward the possible contributions of circular and linear dichroisms as reporting quantities of the presence of chiral molecules in the vicinity of the resonators. These results presented here provide new mechanisms and detection schemes for the sensing of chiral molecules based on optically active achiral metasurfaces. 

\section*{Funding}
This work has been carried out thanks to the support of the ANR (n$^{\circ}$ ANR-18-CE09-0010) and thanks to the support of the Aurora PHC program. BG and SB acknowledge support by the Cluster of Excellence MATISSE funded by the Investissements d'Avenir French Government program. 

\section*{Supporting Information}
Supporting Information file describes: the FEM Method; the FEM calculated Mueller matrix; the antisymmetric part of the logarithm of the calculated Mueller matrix; the symmetric and antisymmetric parts of the logarithms of the measured Mueller matrix; the Mueller matrix measured at normal incidence with the CDOA and estimation of the linear retardance effects on CDAO; plot of the azimuthal variations of the element (2,3) of the measured Mueller matrix; the calculated near-field x-, y- and z-components of the electric field; calculated near-field maps at 0° and 15° of incidence.

\section*{Acknowledgments}
The authors acknowledge the R\'{e}seau des Salles Blanches Parisienne for granting access to the e-beam lithography.

\end{document}